\documentstyle[12pt]{article}

\begin{document}

\def\b#1{\mbox{\footnotesize\boldmath$#1$\normalsize}}

\def\va#1#2#3{{\,{#1}_{#2}\cdots\,{#1}_{#3}}}

\def\vb#1#2#3#4{{\,{#1}_{#3}{#2}_{#3}\cdots\,{#1}_{#4}{#2}_{#4}}}

\def\DG{\mbox{\large$\delta$\normalsize}}

\def\aa{\dot}


\title{THE GAUSS CONSTRAINT IN THE EXTENDED LOOP REPRESENTATION}

\author{\small Cayetano Di Bartolo \\
\small Departamento de F\'{\i}sica, Universidad Sim\'on Bol\'{\i}var,
Apartado 89000,\\
\small \hspace{0.3cm}Caracas 1080-A, Venezuela. \normalsize}

\date{ }
\maketitle

\abstract{The Gauss constraint in the extended loop 
representation for quantum gravity is studied. It is 
shown that there exists a sector of the state space 
that is rigorously gauge invariant without the generic 
convergence issues of the extended holonomies.}

\noindent {\small {\bf pacs:} 11.15.-q, 04.60.-m \normalsize}

\vspace{1cm}

The Ashtekar variables \cite{ash1,ash2} present general 
relativity as a gauge theory. As a consequence, it is 
possible to quantize the theory in the loop 
space \cite{rovsmo88}. In this
representation, the wave functions are loop dependent. All 
the information relevant to the loop is found in its 
'multitangents fields' which can be thought of as loop 
coordinates \cite{extended}. 
The loop representation poses some advantages over the 
connection representation: the Gauss cons\-traint is solved 
in the former, and furthermore, using knot invariant wave 
functions, the diffeomorphism constraint is solved.

Recently, the gravity extended loop representation was  introduced 
\cite{arena,gravext} as a generalization of the loop
representation. In this new representation, the wave function 
domain is the extended loop space. This vector space of 
infinite dimension includes the loop coordinates as a 
particular subset. The extended loop 
representation has several interesting features: it allows to
deal with regularization problems of wave functions which are
inherent to the loop representation; it supplies \cite{remarks}
a new context in which to deal with constraint regula\-rization 
and renormalization, and it supplies new calculation tools 
to handle loop dependent objects.

In this work we will discuss the problem of the gauge 
covariance of the extended representation, some aspects 
of this problem were pointed out and discussed in  
references  \cite{remarks,troy}.
We shall show here that there exists a sector of the state space 
which is gauge invariant. This sector contains most of the wave
functions know in the extended loop representation.

In order to deal with the extended loop representation, it
is useful to define a number of vector spaces.
We define the vector space $\cal E$ of infinite dimension as
the space of all the linear combinations of a 
linearly independent base of vectors $\b\DG_{\aa\nu}$ (the subindex 
${\aa\nu}$ labels the elements of the base),

\begin{equation}
{\b E} \in {\cal E} \leftrightarrow {\b E}=
{\b\DG}_{\aa\nu}E^{\aa\nu} \; .
\label{Espacio1}
\end{equation}

\noindent
The dotted Greek indexes correspond to an ordered set of Greek indexes
$ {\aa \nu } =  \nu_1 \va \nu 2n $
where $n=n({\aa\nu })$ is the number of elements in the set. 
A Greek index has a discrete part and a continuous part,
$ \mu =  (\, a,\,x\,)$ with $a\,\in\, \{1,2,3 \} $ and
$ x\, \in\, R^3 $. We are adopting the generalized Einstein 
convention for repeated indexes

\begin{equation}
\varphi_{\mu}E^{\mu} = \sum^3_{i=1} \int d^3x \varphi_{ix} E^{ix} 
\;\;\; \mbox{ and } \;\;\;
\varphi_{\aa\mu }E^{\aa\mu } = \varphi E + \sum_{n=1} \varphi_{\va \mu 
1n} 
E^{\va \mu 1n} \; .
\end{equation}

\noindent	      
In (\ref{Espacio1}), $E^{\va \nu 1n}$ is the component of range n of 
${\b E}$. The components of the vector ${\b\DG}_{\aa\nu}$ are:

\begin{equation}
\DG^{\aa\mu}_{\aa\nu}\,=\,
\left\{ 
 \begin{array}{ll}
 0 & \mbox{if $\, n(\aa\mu)\,\neq \,n(\aa\nu)$} \\
 1 & \mbox{if $ \,n(\aa\mu)\,= \,n(\aa\nu)=0$} \\
 \mbox{$ \DG^{\mu_1}_{\nu_1}\cdots\DG^{\mu_n}_{\nu_n}$} & 
 \mbox{if $ n=n(\aa\mu)\,= \,n(\aa\nu)\geq 1$}
 \end{array}
\right.
\end{equation}

\noindent
where

\begin{equation}
\DG^{ax}_{by} \, \equiv \delta^a_b \, \delta(x-y) \, .
\end{equation}

\noindent
In the space ${\cal E}$, we introduce the vector product:

\begin{equation}
{\b E_1} \times {\b E_2} \equiv
{\b\DG}_{\aa\mu\,\aa\nu}E^{\aa\mu}_1 E^{\aa\nu}_2 \;. 
\label{producto}
\end{equation}

The extended loop space \cite{extended} ${\cal E}_y $ is the 
vector subspace of elements of ${\cal E}$ that satisfy 
the following differential constraint:

\begin{equation}
\frac{\partial }{\partial x^a} E^{{\aa\alpha}ax{\aa\beta}}=
\DG^{\aa\alpha}_{\aa\nu\, ax}E^{\aa\nu ax \aa\beta } -
\DG^{\aa\beta}_{ax \,\aa\nu}E^{\aa\alpha ax \aa\nu }+
\delta_{x,y} \,\lbrack \DG^{\aa\alpha} E^{\aa\beta } -
\DG^{\aa\beta}E^{\aa\alpha}\rbrack
\label{VD}
\end{equation}

\noindent
where there is a sum over "a" and "y" is the
origin of the extended loop. This equation can be written as
follows (without explicitly indicating the 
$\DG^{\aa\mu}_{\aa\nu}$ quantities):

\begin{equation}
\frac{\partial }{\partial x^a} E^{\vb ax1{i} ax \vb ax{i+1}n}=
\lbrack \delta (x_{i}-x) - \delta (x_{i+1}-x) \rbrack
E^{ \vb ax1n }
\end{equation}

\noindent
where $0\leq i \leq n$ and $x_0\equiv x_{n+1}\equiv y$.
The differential constraint relates the components of range
n+1 and range n of ${\bf E}$, and basically, 
indicates that the completely transverse parts of 
${\bf E}$ are free, i.e., to solve the constraint means to
establish an isomorphism between ${\cal E}_y$ 
and the transverse vector space \cite{extended}. 
It can be shown that the product(\ref{producto}) 
is closed in ${\cal E}_y$. The vectors in ${\cal E}_y$ 
with the positive null range component together with
the product (\ref{producto}) define the extended loop group.

We define ${\cal N}$ as the space constituted by the
linear applications $\varphi$, $\varphi: \cal E \to C\llap{/}$, 
that satisfy

\nobreak
\begin{equation}
\exists \, \mbox{ m integer } / \,
\varphi({\b E})= \varphi_{\aa\mu} E^{\aa\mu} = \sum_{n=0}^{m}
{\varphi}_{\va \mu 1n} E^{\va \mu 1n} \;\;.
\end{equation}

The extended loop representation for a gauge theory or for 
quantum gravity is dual to the connection representation, and
it is formally obtained from the extended loop transform,

\begin{equation}
\varphi({\b E})= \int DA \, \hat{\varphi}(A) \, W_{\b E}[A]
\label{transformada}
\end{equation}

\noindent
where

\begin{equation}
W_{\b E}[A]= Tr[A_{\aa \mu}]E^{\aa \mu}
\label{wilson}
\end{equation}

\noindent
is the Wilson extended functional and $A_{\aa \mu}$ denotes the 
product

\begin{equation}
A_{\vb ax1n} = A_{a_1}(x_1)\cdots A_{a_n}(x_n) \, .
\end{equation}

\noindent
The ${\b E}$ coordinate in (\ref{wilson}) is a vector
in ${\cal E}_y$. However, the origin of extended coordinate 
is not relevant. Due to the trace in (\ref{wilson}), the cyclic 
part of ${\b E}$ is the only one that contributes, this
part loses every notion of the origin; in effect, it 
inherits from (\ref{VD}) the following differential constraint
without origin:

\begin{equation}
V^{\aa\alpha}_{x\,\,\aa\mu}E^{\aa\mu}=
-\frac{\partial }{\partial x^a} E^{(ax{\aa\alpha})_c}+
\lbrack {\b\DG}_{\aa\nu} \, , \, 
{\b\DG}_{ax} \rbrack^{\aa\alpha} \; E^{(ax {\aa\nu})_c}=0
\label{VDC}
\end{equation}

\noindent
with

\begin{equation}
V^{\aa\alpha}_{x\,\,\aa\mu}\equiv
-\frac{\partial }{\partial x^a} {\b\DG}^{(ax{\aa\alpha})_c}_{\aa\mu}+
\lbrack {\b\DG}_{\aa\nu} \, , \, 
{\b\DG}_{ax} \rbrack^{\aa\alpha} \; {\b\DG}^{(ax {\aa\nu})_c}_{\aa\mu}
\label{VDC2}
\end{equation}

\noindent
where there is a sum in 'a' and no integration in 'x',
the notation $(\,)_c$ meaning the sum of all the cyclic 
permutations of indices
and the commutator is formed with the product '$\times$'. 
Formally, the differential constraint guarantees the gauge
invariance of $W_{\b E}[A]$ and of the theory. However, the
functional (\ref{wilson}) is not well defined for all the
elements of the extended loop space. There exist pairs of 
connections and vectors of the extended loop space for 
which the infinite sum in (\ref{wilson}) does not converge;
but we shall see that there is a sector in the state space
where this problem is not important.

Under the infinitesimal gauge transformation,

\begin{equation}
(\delta A_{ax})= \Lambda_{x,a} + A_{ax}\Lambda_x-\Lambda_x A_{ax}
\label{calibre}
\end{equation}

\noindent
the quantity $A_{\aa\mu}$ transforms according to

\begin{equation}
\delta A_{\va \mu 1n}= \sum^n_{h=1} 
A_{\va\mu 1{h-1}} \, (\delta A_{\mu_h}) \, A_{\va\mu {h+1}n }
=
\delta_{\va \mu 1n}^{\aa\alpha\,ax\,\aa\beta}
A_{\aa\alpha}(\delta A_{ax})A_{\aa\beta} \,.
\end{equation}

\noindent
Its trace satisfies

\begin{equation}
\delta Tr[A_{\aa\mu}] = \delta^{\aa\alpha ax \aa\beta}_{\aa\mu}
Tr\lbrack A_{\aa\beta \aa\alpha} (\delta A_{ax}) \rbrack =
\delta^{\aa\alpha ax \aa\beta}_{\aa\mu} 
\delta^{\aa\nu}_{\aa\beta \aa\alpha}
Tr\lbrack A_{\aa\nu} (\delta A_{ax}) \rbrack \, .
\end{equation}

\noindent
Using in this expression, the identity

\begin{equation}
\delta^{{\aa\alpha}\, ax \, {\aa\beta}}_{\aa\mu}
\delta^{\aa\nu}_{\aa\beta\,\aa\alpha}
= \delta^{ (ax \, {\aa\nu})_c}_{\aa\mu}
\end{equation}

\noindent
and replacing $\delta A_{ax}$ from (\ref{calibre}), we get

\begin{equation}
\delta Tr[A_{\aa\mu}] = 
Tr\lbrack -\Lambda_x A_{\aa\nu} \frac{\partial }{\partial x^a}+
\Lambda_x A_{\aa\alpha} (\delta^{\aa\alpha}_{\aa\nu ax}-  
\delta^{\aa\alpha}_{ax \aa\nu})\rbrack 
\delta^{(ax \, \aa\nu)_c}_{\aa\mu}
=
Tr\lbrack \Lambda_x A_{\aa\alpha}\rbrack\, V^{\aa\alpha}_{x\aa\mu}
\;.
\label{deltatraza}
\end{equation}

We introduce the gauge dependent quantities

\begin{equation}
\varphi^{\va I1n}_{\va \mu 1n} \equiv \int DA \, \hat{\varphi}(A) \,
A^{I_1}_{\mu_1}\cdots A^{I_n}_{\mu_n}
\end{equation}

\noindent
and

\begin{equation}
\varphi_{\va \mu 1n} \equiv \varphi^{\va I1n}_{\va \mu 1n}
Tr(T^{I_1} \cdots T^{I_n}) = 
\int DA \, \hat{\varphi}(A) \, Tr(A_{\mu_1}\cdots A_{\mu_n})
\end{equation}

\noindent
where the $T^I$ are the gauge group generators. Because of 
(\ref{deltatraza}), $\varphi_{\aa\mu}$ transforms according to

\begin{equation}
\delta\varphi_{\aa\mu}= \int DA \,
Tr\lbrack \Lambda_x A_{\aa\alpha}\rbrack\, V^{\aa\alpha}_{x\aa\mu}
\label{transfo1}
\end{equation}

\noindent
under the gauge change (\ref{calibre}). We shall suppose now that
$\hat{\varphi}(A)$ is such that in some gauge G the following
"cut condition" is satisfied:

\begin{equation}
\exists \;M\;\; / \; \varphi^{\va I1n}_{\va \mu 1n}=0 
\;\; \forall n \,  > \, M \, .
\label{corte}
\end{equation}

\noindent
This implies that in the gauge G the function

\begin{equation}
\varphi({\b E})= \int DA \, \hat{\varphi}(A) \, 
Tr[A_{\aa\mu}] E^{\aa\mu}
= \varphi_{\aa\mu} E^{\aa\mu} \; ,
\end{equation}

\noindent
is not dependent on the components of range $n(\aa\mu)>M$ 
in ${\b E}$. Next, let us show that $\varphi({\b E})$ 
is invariant under infinitesimal gauge changes. We
shall only take gauge transformations for which $\Lambda_x$ 
can be expanded in a 'power series' in the connection,

\begin{equation}
\Lambda_x=\lambda_x + \lambda^{ J,\mu}_x 
A^J_\mu + \cdots \; .
\label{serie}
\end{equation}

From (\ref{transfo1}) we have 

\begin{equation}
\delta \varphi({\b E})=
\int DA \,
Tr\lbrack \Lambda_x A_{\aa\alpha}\rbrack\, V^{\aa\alpha}_{x\aa\mu}
E^{\aa\mu}
\label{transfo2}
\end{equation}

\noindent
In this expression, as a consequence of the "cut condition"
the sum in $\aa\alpha$ is finite, \linebreak
( $Max\,\,n(\aa\alpha)=M$ ), and
from the definition of $V^{\aa\alpha}_{x\aa\mu}$, we have
that $Max\,\,n(\aa\mu)=M+1$, i.e., the sum in (\ref{transfo2})
involves a finite number of ranges of ${\b E}$, and because
(\ref{VDC}), we have

\begin{equation}
\delta \varphi({\b E})=0
\end{equation}

The functionals 
$\varphi({\b E})=\varphi_{\aa\mu} E^{\aa\mu}$, 
which satisfy the "cut condition", are gauge invariant, and
belong to ${\cal N}$. These functionals satisfy Mandelstam
identities which reflect the specific structure of the 
gauge group \cite{gravext}. In effect, for SU(2), the 
Mandelstam identities  -which relates group generator 
product traces- lead to the following relations:

\begin{eqnarray} 
\varphi_{\aa\mu\aa\nu} =&& \varphi_{\aa\nu \aa\mu} \nonumber \\
\varphi_{\aa \mu } =&& \varphi_{ \overline{\aa\mu}}  
\label{mandelstam}\\
\varphi_{\aa\mu \aa\nu \aa\gamma} + 
\varphi_{\aa\mu \aa\nu \overline{\aa\gamma}}
=&& 
\varphi_{\aa\nu \aa\mu \aa\gamma} + 
\varphi_{\aa\nu \aa\mu \overline{\aa\gamma}}
\nonumber
\end{eqnarray}

\noindent
where to overline a set of Greek indexes means to invert
the order and the multiplication by a sign, as follows:

\begin{equation} 
\delta^{\overline{\va\mu 1n}}_{\aa \nu}\equiv
\delta^{\va\mu 1n}_{\overline{\aa \nu}}\equiv
(-1)^n \delta^{\va\mu n1}_{\aa \nu} \, .
\end{equation}

\noindent
The set of functionals that belong to 
${\cal N}$ and that satisfy the identities (\ref{mandelstam})
contains a sector, gauge invariant, of the state space of the
quantum gravity extended loop representation.

In quantum gravity, the wave functions, 
$\varphi({\b E})= \varphi_{\aa\mu} E^{\aa\mu}$, must be invariant
under diffeomorphism. This implies that by evaluating 
$\varphi({\b E})$ in the coordinate of a loop $\Gamma$,
${\b E} = {\b X}(\Gamma)$, a knot invariant is obtained.
Reciprocally, if we have a knot invariant of the form

\begin{equation} 
\Psi(\Gamma) = \Psi'({\b X(\Gamma)}) = \,
\sum^M_{n=0} \Psi_{\va\mu 1n} {\b X(\Gamma)}^{\va\mu 1n}
\, ,
\label{nudo}
\end{equation}

\noindent
the function $\Psi'({\b E})$ belongs to the state space in 
the extended loop representation. If the perturbative 
expansion of the expectation value of the Wilson functional 
in the Chern-Simons theory is considered, every one of
orders of the expansion has the form \cite{a3} (\ref{nudo}).
An example of a knot invariant which is not a finite linear 
combination of the form (\ref{nudo}) is the exponential of 
the Gauss invariant. However, there exist an infinite
sequence of functionals in ${\cal N}$ which tend to the 
exponential of the Gauss invariant.

The following is a possible generalization of the 
exponential of the Gauss invariant in the extended loop
representation:

\begin{equation} 
\exp^{*}({\b E}) \equiv \lim_{M\to\infty} F_M({\b E})
\label{ex}
\end{equation}

\noindent
where

\begin{equation} 
F_M({\b E}) \equiv \sum^M_{n=0} \frac{a^n }{2^n n!}
g_{\mu_1\,\nu_1} \cdots g_{\mu_n\,\nu_n}
E^{(\mu_1\,\nu_1 \cdots \mu_n\,\nu_n)_S}
\label{FM}
\end{equation}

\noindent
and $g_{ax \, by} \equiv - \varepsilon_{abc} 
\partial_c \nabla^{-2} \delta (x-y)$
is the propagator of the Chern-Simons theory. In (\ref{FM}),
the subscript 'S' indicates a sum over all the permutations
of the Greek indexes. The functions $F_M(\b E)$ belong
to ${\cal N}$, satisfy the Mandelstam identities
(\ref{mandelstam}) and, it can be shown that, they are
diffeomorphism invariant.\footnote{For each $n$, 
$E^{\va \nu 1n}$ is a vector density of weight one in 
each argument $\nu=(ax)$.}  By evaluating
(\ref{ex}) in loop coordinates, we get

\begin{equation} 
\exp^*[{\b X(\Gamma)}] = \exp [a \,\rho(\Gamma)]
\label{gauss}
\end{equation}

\noindent
where $\rho(\Gamma)$ is the Gauss self-linking number of
the loop $\Gamma$. In order to obtain (\ref{gauss}), the
following algebraic constraint, satisfied by the loop
coordinate, was used,

\begin{equation} 
[\b X(\Gamma)]^{{\aa\beta}_1\nu_1 \cdots {\aa\beta}_n\nu_n
{\aa\beta}_{n+1} }
 \delta^{\va \mu 1m}_{\va {\aa\beta} 1{n+1}}
=
[\b X(\Gamma)]^{\va \mu 1m}\, [\b X(\Gamma)]^{\va \nu 1n}
\;.
\label{VA}
\end{equation}

As shown in reference  \cite{GaPu},
$\exp [-3\Lambda \,\rho(\Gamma)/2 ]$
is a formal solution of gravity with cosmological constant
in the loop representation. It is expected that its
extension (\ref{ex}) is the corresponding solution in the
extended loop representation. However, $\exp^*$ does not
converge for every vector $\b E$. A possible solution to
this problem would be to limit the extended loop space to
those vectors that satisfy an algebraic constraint equal to 
(\ref{VA}). In this case, (\ref{ex}) converges to

\begin{equation} 
\exp^*[{\b E}] = 
\exp [\frac{a}{2} g_{\mu_1\mu_2}\, E^{\mu_1}\,E^{\mu_2}] \,.
\end{equation}

The problem of the appropriate definition of $\exp^*$ and 
the action of gravity constraints on this functional is 
still under study. It is interesting to point out that 
there are also problems in the lattice, where it has not 
been possible to define an exponential of the Gauss number
which simultaneously satisfies the Hamilton and diffeomorphism
constraint \cite{privada}.

\vspace{0.5cm}
We wish especially to thank Rodolfo Gambini for his
critical comments.

\end{document}